\documentstyle[bo99,epsfig]{article}

\newcommand{\asca}{{\it ASCA}}
\newcommand{\rosat}{{\it ROSAT}}
\newcommand{\sax}{{\it BeppoSAX}}
\newcommand{\einstein}{{\it Einstein}}
\newcommand{\ginga}{{\it Ginga}}
\newcommand{\heao}{{\it HEAO1}}

\newcommand{\logn}{Log $N$ - Log $S$ relation}

\newcommand{\etal}{{\it et al.}}

\newcommand{\erg}{erg s$^{-1}$}
\newcommand{\ergs}{erg s$^{-1}$ cm$^{-2}$}

\newcommand{\de}{deg$^2$}

\title{Results from X-ray Surveys with ASCA }

\author{Yoshihiro~Ueda}
\affil{Institute of Space and Astronautical Science, Kanagawa 229-8510, Japan}                                                
\begin{document}

\maketitle

\begin{abstract}

We present main results from X-ray surveys performed with \asca ,
focusing on the \asca\ Large Sky Survey (LSS), the Lockman Hole deep
survey, and the \asca\ Medium Sensitivity Survey (AMSS or the GIS
catalog project). The \logn s, spectral properties of sources, and
results of optical identification are summarized. We discuss
implications of these results for the origin of the CXB.

\end{abstract}

\section{Introduction}

Understanding the origin of the Cosmic X-ray Background (CXB or XRB)
and cosmological evolution of X-ray extragalactic populations is one 
of the main goals of X-ray astronomy. In the soft X-ray band, the
\rosat\ satellite resolved 80\% of the 0.5--2 keV CXB into
individual sources (Hasinger \etal\ 1998) and optical identification
revealed that the major population is type-I AGNs (Schmidt \etal\
1998). Because of the technical difficulties, imaging sky surveys in
the hard X-ray band (above 2 keV), where the bulk of the CXB energy
arises, were not available until the launch of \asca . The sensitivity
limits achieved by previous mission such as \heao\ (Piccinotti \etal\
1982) and \ginga\ (Kondo \etal\ 1991) are at most $\sim 10^{-11}$
\ergs\ (2--10 keV), and the sources observed by them only account for
3\% of the CXB intensity in the 2--10 keV band. In particular, there
is a big puzzle on the CXB origin, called the ``spectral paradox'':
bright AGNs observed with \heao, {\it EXOSAT} and \ginga\ have spectra
with an average photon index of $\Gamma$ = 1.7$-$1.9 (e.g., Williams
\etal\ 1992), which is significantly softer than that of the CXB
itself ($\Gamma \simeq $ 1.4; e.g., Gendreau \etal\ 1995). 
Furthermore, the broad band properties of sources at fluxes from
$\sim10^{-11}$ to $\sim 10^{-13}$ \ergs\ (2--10 keV) are somewhat
puzzling according to previous studies. The extragalactic source counts
in the soft band (0.3--3.5 keV) obtained by \einstein\ Extended Medium
Sensitivity Survey (EMSS; Gioia \etal\ 1990) is about 2--3 times
smaller than that in the hard band (2--10 keV) obtained by the \ginga\
fluctuation analysis (Butcher
\etal\ 1997) when we assume a power-law photon index of 1.7.

The \asca\ satellite (Tanaka, Inoue, \& Holt 1994), launched in 1993
February, was expected to change this situation. It is the first
imaging satellite capable of study of the X-ray band above 2 keV with
a sensitivity up to several $ 10^{-14}$\ergs\ (2--10 keV) and covers
the wide energy band from 0.5 to 10 keV, which allows us to directly
compare results of the energy bands below and above 2 keV with single
detectors, hence accompanied with much less uncertainties than
previous studies. By taking these advantages, several X-ray surveys
have been performed with \asca\ to reveal the nature of hard X-ray
populations: the \asca\ Large Sky Survey (LSS; Ueda \etal\ 1998), the \asca\
Deep Sky Survey (DSS; Ogasaka \etal\ 1998; Ishisaki \etal\ 1999 for the
Lockman Hole), the \asca\ Medium-Sensitivity Survey (AMSS or the GIS
catalog project: Ueda \etal\ 1997, Takahashi \etal\ 1998, Ueda \etal\
1999b; see also Cagnoni, Della Ceca, \& Maccacaro 1998 and Della Ceca
\etal\ 1999), a survey of \rosat\ deep fields (Georgantopoulos \etal\
1997; Boyle \etal\ 1998), and so on. The sensitivity limits and survey
area are summarized in Table~1. In this paper, we present main results
of the \asca\ surveys, focusing on the LSS (\S~2), the Lockman Hole
deep survey (\S~3), and the AMSS (\S~4). In \S~5, we summarize these
results and discuss their implications for the origin of the CXB.

\begin{table}
\begin{small}
\begin{center}
\caption[]{Summary of \asca\ Surveys}
\begin{tabular}{lll}
\hline\hline
Survey Project& Area & Sensitivity (2--10 keV)\\
      & (deg$^2)$      & (\ergs )\\
\hline
Large Sky Survey (LSS)& 7.0 & 1.5$\times 10^{-13}$\\
Deep Sky Survey (DSS)& 0.3 & 4$\times 10^{-14}$ \\
Lockman Hole Deep Survey& 0.2 & 4$\times 10^{-14}$\\
Survey of deep \rosat\ fields& 1.0 & 5$\times 10^{-14}$\\
\asca\ Medium-Sensitivity Survey (AMSS) & 110 & 7$\times 10^{-14}$\\
\hline
\end{tabular}
\end{center}
\end{small}
\end{table}

\section{The Large Sky Survey}

\subsection{X-ray Data}

The survey field of the \asca\ Large Sky Survey (LSS; Ueda \etal , 1998)
is a continuous region near the north Galactic pole, centered at
$RA$(2000) = 13$^{\rm h}$14$^{\rm m}$, $DEC$(2000) = 31$^\circ 30'$. 
Seventy-six pointings have been made over several periods from Dec.\
1993 to Jul.\ 1995. The total sky area observed with the GIS and SIS
amounts to 7.0 \de\ and 5.4 \de\ with the mean exposure time of 56
ksec (sum of GIS2 and GIS3) and 23 ksec (sum of SIS0 and SIS1),
respectively. From independent surveys in the total (0.7--7 keV), hard
(2--10 keV), and soft (0.7--2 keV) bands, 107 sources are detected
with sensitivity limits of $6\times 10^{-14}$, $1\times 10^{-13}$, and
$2\times 10^{-14}$ \ergs\ , respectively. The \logn s derived from the
LSS are summarized in Ueda \etal\ (1999a) together with a complete
X-ray source list. At these flux limits, 30($\pm$3)\% of the CXB in
the 0.7--7 keV band and 23($\pm$3)\% in the 2--10 keV band have been
resolved into discrete sources. The 2--10 keV \logn\ combined with 
the AMSS result (\S 4) is plotted in Figure~3. 

The spectral properties of the LSS sources suggest that contribution
of sources with hard energy spectra become significant at a flux of
$\sim 10^{-13}$ \ergs\ (2--10 keV), which are different from the major
population in the soft band. The average 2--10 keV photon index is
1.49$\pm$0.10 (1$\sigma$ statistical error in the mean value) for 36
sources detected in the 2--10 keV band with fluxes below
$4\times10^{-13}$ \ergs , whereas it is 1.85$\pm$0.22 for 64 sources
detected in the 0.7--2 keV band with fluxes below $3\times10^{-13}$
\ergs . The average spectrum of 74 sources detected in the 0.7--7 keV
band with fluxes below $2\times10^{-13}$ shows a photon index of
1.63$\pm$0.07 in the 0.7--10 keV range: this index is consistent with
the comparison of source counts between the hard and the soft band.

To investigate the X-ray spectra of these hard sources, we made deep
follow-up observations with \asca\ for the five hardest sources in the
LSS, selected by the apparent 0.7--10 keV photon index from the source
list excluding very faint sources. The results are summarized in Ueda
\etal\ (1999c); see also Sakano \etal\ (1998) and Akiyama \etal\ (1998)
for AX~J131501+3141, the hardest source in the LSS. Three sources in
this sample are optically identified as narrow-line AGNs and one is a
weak broad-line AGN by Akiyama \etal\ (2000); one is not identified
yet. We found that spectra of these sources are most likely subject to
intrinsic absorption at the source redshift with column densities of
$N_{\rm H} = 10^{22} \sim 10^{23}$ cm$^{-2}$.

\subsection{Optical Identification}

Akiyama \etal\ (2000) summarize the results of optical identification
for a sub-sample of the LSS sources, consisting of 34 sources detected
in the 2--7 keV band with the SIS. The major advantage
of this sample compared with other \asca\ surveys is good position
accuracy; it is 0.6 arcmin in 90\% radius from the \asca\ data alone,
thanks to superior positional resolution of the SIS. To improve the
position accuracy further, we made follow-up observations with {\it
ROSAT} HRI over a part of the LSS field in Dec.\ 1997. Optical
spectroscopic observations were made using the University of Hawaii
88$''$ telescope, the Calar Alto 3.5m telescope, and the Kitt Peak
National Observatories Mayall 4m and 2.1m telescopes.

Out of the 34 sources, 30 are identified as AGNs, 2 are clusters of
galaxies, 1 is a Galactic star, and only 1 object remains
unidentified. The identification as AGNs is based on existence of a
broad emission line or the line ratios of narrow emission lines
([NII]6583$\AA$/H$\alpha$ and/or [OIII]5007$\AA$/H$\beta$); see
Akiyama \etal\ 2000 and references therein.  Figure~1(a) shows the
correlation between the redshift and the apparent photon index in the
0.7--10 keV range, which is obtained from a spectral fit assuming no
intrinsic absorption, for the identified objects. The 5 sources that
have an apparent photon index smaller than 1.0 are identified as 4
narrow-line AGNs and 1 weak broad-line AGN, all are located at
redshift smaller than 0.5. On the other hand, X-ray spectra of the
other AGNs are consistent with those of nearby type 1 Seyfert
galaxies. Four high redshift broad-line AGNs show somewhat apparently
hard spectra with an apparent photon index of $1.3\pm0.3$, although it
may be still marginal due to the limited statistics.

To avoid complexity in classifying the AGNs by the optical spectra, we
divide the identified AGNs into two using the X-ray data: the
``absorbed'' AGNs which show intrinsic absorption with a column
density of $N_{\rm H} > 10^{22}$ cm$^{-2}$ and the ``less-absorbed''
AGNs with $N_{\rm H} < 10^{22}$. Correcting the {\it flux} sensitivity
for different X-ray spectra, we found the contribution of the
absorbed AGNs is almost comparable to that of less-absorbed
AGNs in the 2--10 keV source counts at a flux limit of
$2\times10^{-13}$ \ergs .  Figure~1(b) shows the correlation between
the redshift and the 2--10 keV luminosity of the identified AGNs. The
redshift distribution of the 5 absorbed AGNs is concentrated at
$z<0.5$, which contrasts to the presence of 15 less-absorbed AGNs at
$z>0.5$. This suggests a deficiency of AGNs with column densities of
$N_{\rm H} = 10^{22-23}$ at $z$ = 0.5--2, or in the X-ray luminosity
range larger than $10^{44}$ \erg , or both. Note that if the 4
broad-line AGNs with hard spectra have intrinsic absorption instead of
other hardening mechanism such as Compton reflection, it could
complement this deficiency.

\bigskip
\bigskip
\begin{figure}[htp]
\centerline{\psfig{file=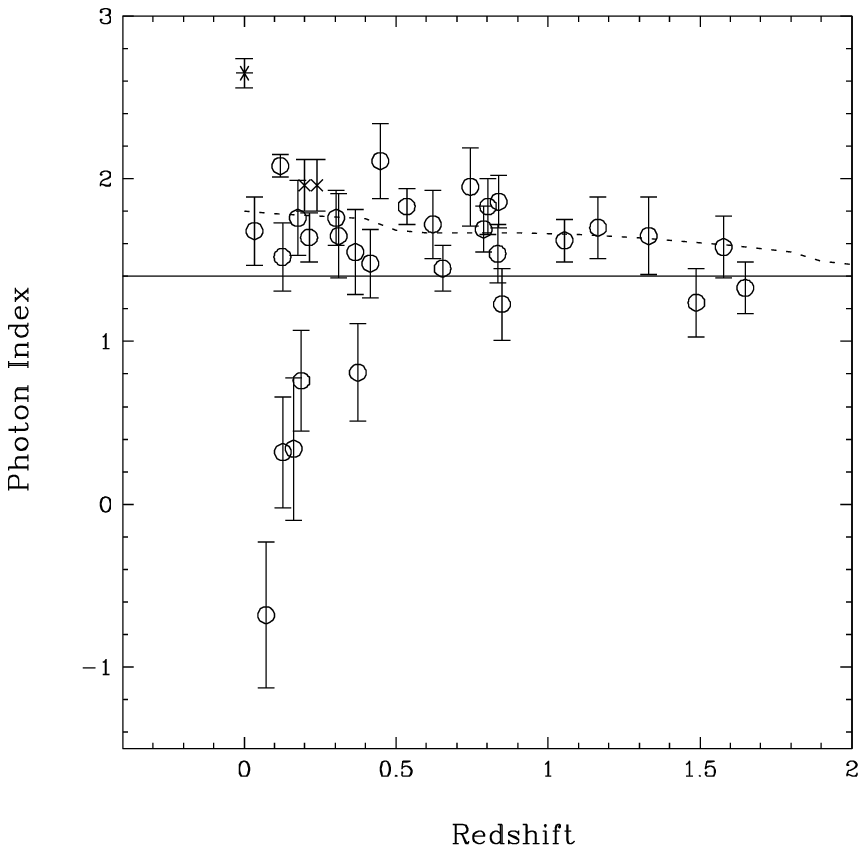, width=6cm}\hspace{1cm}\psfig{file=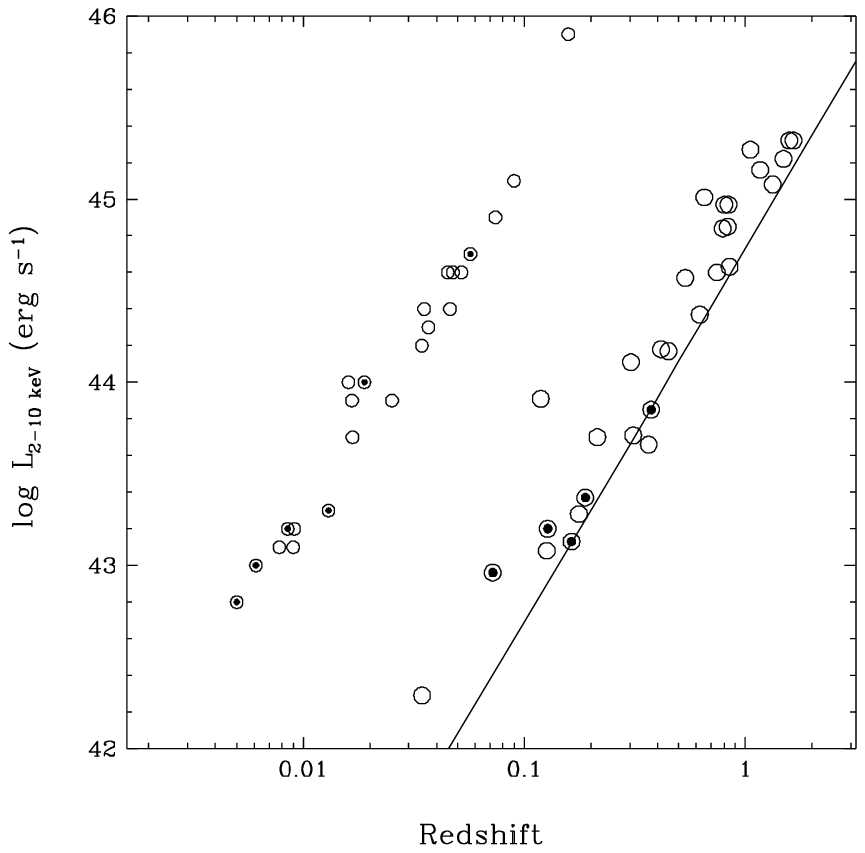, width=6cm}}
\caption[]{(a) left: the correlation between the redshift and 
the apparent 0.7--10 keV photon index for the identified objects in the 
LSS (Akiyama \etal\ 2000). The open circles, 
crosses, and asterisk represent AGNs, clusters of
galaxies, and a Galactic star, respectively.
The dotted curve shows the expected apparent photon index in the
observed 0.7--10 keV band as a function of redshift, for a typical spectrum of 
type-1 Seyfert galaxies with a Compton reflection
component.
(b) right: The 2--10 keV luminosity versus redshift diagram
for the LSS AGNs (with large open circles, Akiyama \etal\ 2000), and for 
the \heao A2 AGNs (with small marks, Piccinotti \etal\ 1982). The
``absorbed'' AGNs are plotted with dots. Lines indicate 
detection limits of the LSS for a source with an photon index of 1.7
with no intrinsic absorption.
}
\end{figure}

\section{The Lockman Hole Deep Survey}

Deep surveys were performed with \asca\ over several fields (Ogasaka
\etal\ 1998), although optical identification is more difficult than
the LSS because of faint flux levels and source confusion problem.  To
overcome this difficulty, we have been conducting a deep survey of the
Lockman Hole, where the \rosat\ deep survey was performed (Hasinger
\etal\ 1998). The advantage of selecting this field is that we already
have a complete soft X-ray source catalog down to a flux limit of
$5.5\times10^{-15}$ \ergs\ (0.5--2 keV), most of which have been
optically identified (Schmidt \etal\ 1998). In addition, we utilized an
X-ray source list at even fainter flux limits (G.~Hasinger, private
communication). Since the flux limits of the \rosat\ surveys are
extremely low, we can expect most of \asca\ sources could have \rosat\
counterparts within reasonable range of spectral hardness. Utilizing
positions of the \rosat\ sources, we can determine the hard-band flux
for individual sources, which would otherwise have been difficult to
separate, to a flux limit of $3\times10^{-14}$ \ergs\ (2--10 keV). 
Preliminary results are reported in Ishisaki \etal\ (1999).
 
Up to present, we have made 3 pointings in the direction of the
Lockman Hole with \asca\ on 1993 May 22--23, 1997 April 29--30, and
1998 November 27 for a net exposure of 63 ksec (average of the 8 SIS
chips), 64 ksec, and 62 ksec, respectively. The pointing positions
were arranged so that the superposed image of the SIS field of views
(FOVs) covers the PSPC and HRI FOVs as much as possible. We here used
only the SIS data considering its superior positional resolution. 
Analysis was made through the 2-dimensional maximum-likelihood fitting
to a raw, superposed image in photon counts space in the sky
coordinates, with a model consisting of source peaks (point spread
functions) and the background. As a first step, we put sources into
the model at the positions of the \rosat\ catalogs. Then, after
checking the residual image of the fit, we added remaining peaks that
were missing in the \rosat\ catalogs. Thus, we determined the
significance and flux of each source in three energy bands, 0.7--7
keV, 2--7 keV, and 0.7--2 keV, including new sources detected with
\asca . We corrected for the degradation of detection efficiency
caused by the radiation damage using the CXB intensity. Note that the
\asca\ sensitivity limits strongly depends on position due to the
multiple pointings and the vignetting of the XRT.

We detected 27 sources altogether with significances higher than
3.5$\sigma$ in either of the three survey bands. Two sources were
newly detected with \asca . One object is a variable source having a
0.7--7 keV photon index of about 1.7, which was very faint during the
\rosat\ observations. The other shows a very hard spectrum and is
detected only in the 2--7 keV band. In the combined SIS FOVs, 43
sources out of 50 sources in the Schmidt \etal\ (1998) catalog are
located. Identification of \asca\ sources using the \rosat\ catalog is
summarized in Table~2. Since the number of sources detected in the
2--7 keV band is limited due to poor photon statistics, we here use
the results for 25 sources detected in the 0.7--7 keV band for
comparison with the \rosat\ survey. Four unidentified
sources in the \asca\ survey have \rosat\ counterparts in the deeper
X-ray source catalog (G.~Hasinger, private communication) and
remaining one is the variable source detected only with \asca . For
AGNs identified by Schmidt \etal\ (1998), we divided them into two
according to their optical spectra: (1) type-1 AGNs, corresponding to
either of class a, b, or c, showing broad emission lines, and (2)
type-2 AGN, class d or e, showing only narrow emission lines. As
noticed from the table, 6 out of 7 type-2 AGNs were detected, whereas
only half of the 26 type-1 AGNs were detected with \asca , which
covers much harder band than the \rosat . This suggests that
contribution of type-2 AGNs are more dominant in higher energy bands
than in the soft band at similar flux levels.

\begin{table}
\begin{small}
\caption[]{Summary of optical identification of the \asca\ Lockman Hole deep survey by the \rosat\ catalog (Schmidt \etal\ 1998)}
\begin{center}
\begin{tabular}{lcc}
\hline\hline
Population& \rosat\ & \asca\ (0.7--7 keV)\\\hline 
Total & 43 & 25 \\\hline
Type-1 AGN (a-c) & 26& 13 \\
Type-2 AGN (d-e) & 7 & 6 \\
Group/Galaxies& 3& 0 \\
Star & 3 & 1 \\
Unidentified & 4 &4+1\\
\hline
\end{tabular}
\end{center}
\end{small}
\end{table}

\section{The \asca\ Medium-Sensitivity Survey}

Because these surveys are limited in sky coverage, the sample size is
not sufficient to obtain a self-consistent picture about the evolution
of the sources over the wide fluxes, from $\sim10^{-11}$ \ergs\ (2--10
keV) which is the sensitivity limit of \heao\ A2 (Piccinotti
\etal\ 1982), down to $\sim 10^{-13}$ \ergs\ (2--10 keV), 
that of \asca\ . To complement these shortcomings, we have been
working on the project called the ``\asca\ Medium Sensitivity Survey
(AMSS)'', or the GIS catalog project. In the project, we utilize the
GIS data from the fields that have become publicly available to search
for serendipitous sources. The large field of view and the
low-background characteristics make the GIS instrument ideal for this
purpose.

Main results from the AMSS are reported in Ueda \etal\ (1999b), which
were obtained from selected GIS fields of $|b|> 20^\circ$ observed
from 1993 to 1996, covering the total sky area of 106 deg$^{-2}$. The
sample contains 714 serendipitous sources, of which 696, 323, and 438
sources are detected in the 0.7--7 keV (total), 2--10 keV (hard), and
0.7--2 keV (soft) band, respectively. This is currently the largest
X-ray sample covering the 0.7--10 keV band. Figure~2(a) shows the
correlation between the 0.7--7 keV flux and the hardness ratio between
the 2--10 keV and 0.7--2 keV count rates. We also plot the average
hardness ratio in several flux ranges, separated by the dashed curves,
with crosses. It is clearly seen that the average spectrum becomes harder
with a decreasing flux: the corresponding photon index (assuming a
power law over the 0.7--10 keV band with no absorption) changes from
2.1 at the flux of $\sim 10^{-11}$ \ergs\ to 1.6 at $\sim 10^{-13}$
\ergs\ (0.7--7 keV). Similar hardening are also reported in the 2--10 keV range 
by Della Ceca \etal\ (1999) using 60 serendipitous sources. 
Figure~2(b) shows the integral \logn s in the 0.7--7 keV survey band
for the soft source sample, consisting of sources with an apparent
0.7--10 keV photon index larger than 1.7, and for the hard source
sample, with an index smaller than 1.7. This demonstrates that sources
with hard energy spectra in the 0.7--10 keV range are rapidly
increasing with decreasing fluxes, compared with softer sources. 

\begin{figure}
\centerline{\psfig{file=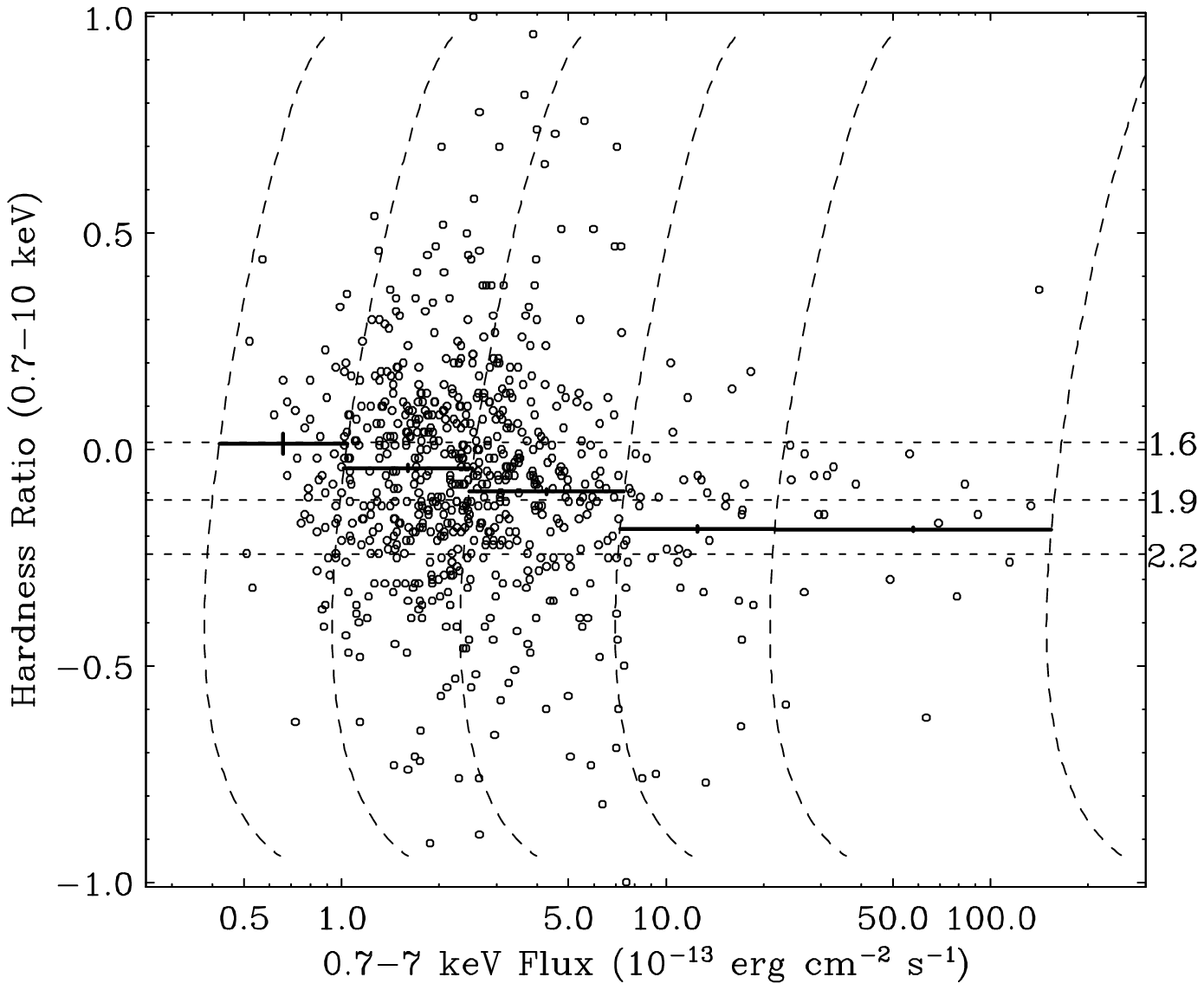, width=6cm}\hspace{0.5cm}\psfig{file=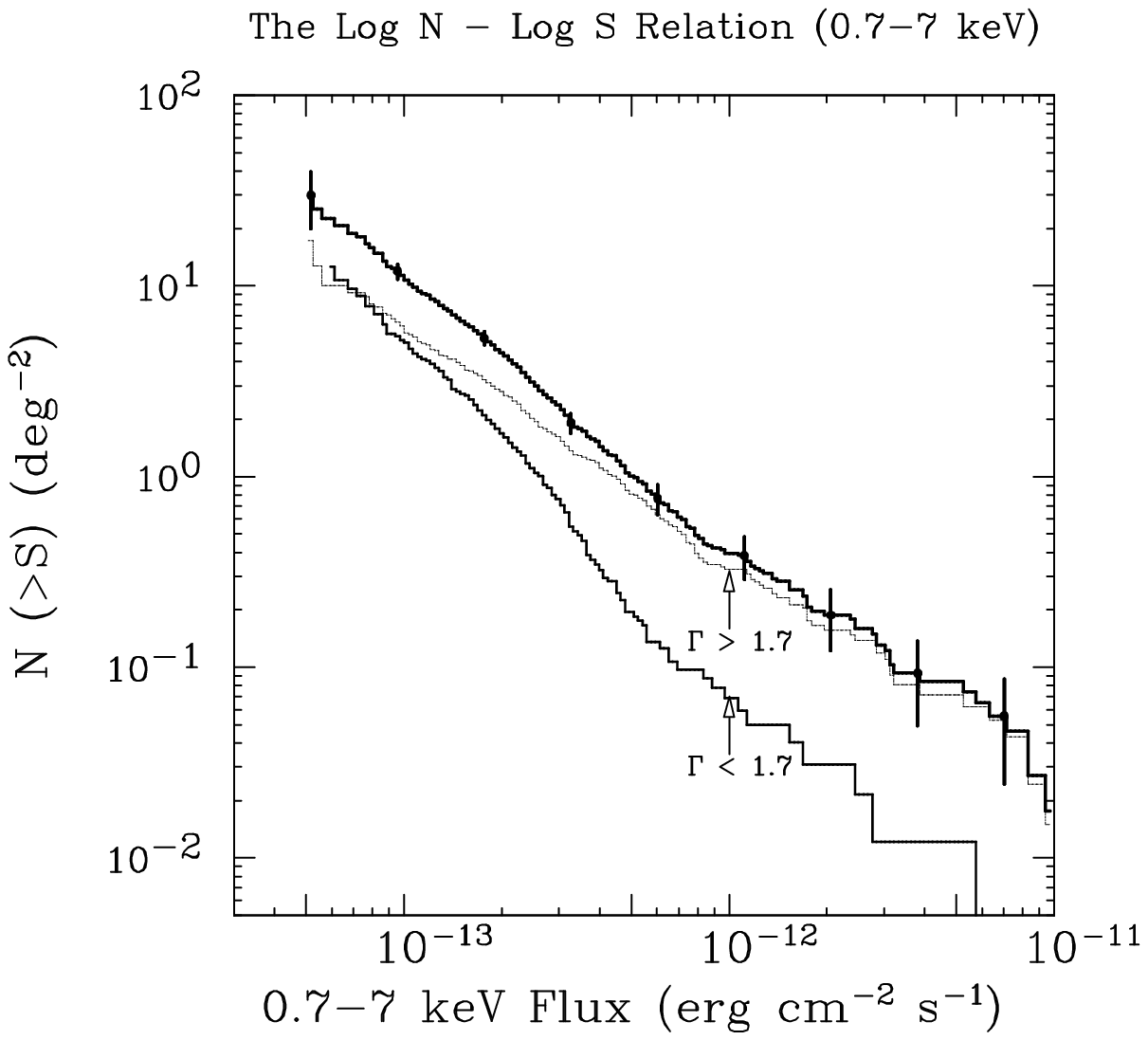, width=6.5cm}}
\caption[]{(a) left: 
The correlation between the 0.7--7 keV flux and the hardness ratio
between the 0.7--2 keV and 2--10 keV count rates for sources detected
in the 0.7--7 keV survey in the AMSS sample (Ueda \etal\ 1999b).
The crosses show the
average hardness ratios (with 1$\sigma$ errors in the mean value) in
the flux bin separated by the the dashed curves, at which the count
rate hence the sensitivity limit is constant. The dotted lines
represent the hardness ratios corresponding to a photon index of 1.6,
1.9, and 2.2 assuming a power law spectrum. 
(b) right: 
The integral \logn s in the 0.7--7 keV survey band, derived from the
AMSS sample. The medium-thickness curve represents the result for the
hard source sample, consisting of sources with an apparent 0.7--10 keV
photon index $\Gamma$ smaller than 1.7, the thin curve represents that
for the soft source sample (with $\Gamma$ larger than 1.7), and the
thick curve represents the sum. The 90\% statistical errors in source
counts are indicated by horizontal bars at several data points.
}
\end{figure}

\section{Summary}

The \asca\ surveys have brought a clear, self-consistent picture about
statistical properties of sources that constitute about 30\% of the
CXB in the broad energy band of 0.7--10 keV. Figure~3 summarizes the
2--10 keV \logn\ obtained from the \asca\ surveys together with the
results from previous missions. The direct source counts from combined
results of the LSS (Ueda \etal\ 1999b) and the AMSS (Ueda \etal\
1999c; these contain the data used by Cagnoni, Della Ceca, \&
Maccacaro 1998) give the tightest constraints so far over a wide flux
range from $\sim 10^{-11}$ to $\sim 7\times10^{-14}$ \ergs : $N(>S)$ =
16.8$\pm$7.2 (90\% statistical error), 11.43$\pm$2.4, 3.76$\pm$0.42,
1.08$\pm$0.17, and 0.33$\pm$0.09 deg$^{-2}$, at $S$ =
$7.4\times10^{-14}$, $1.0\times10^{-13}$, $2.0\times10^{-13}$,
$4.0\times10^{-13}$, and $1.0\times10^{-12}$
\ergs , respectively. The DSS gives a direct source counts at the
faintest flux, $3.8\times10^{-14}$ \ergs (Ogasaka
\etal\ 1998), whereas the fluctuation analysis of deep SIS fields constrains
the \logn\ at fluxes down to $1.5\times10^{-14}$ (Gendreau, Barcons, \&
Fabian 1998). As seen from the figure, the \asca\ direct source counts
smoothly connect the two regions constrained by the \ginga\ and
\asca\ fluctuation analysis.

The AMSS/LSS results demonstrate that the average spectrum of X-ray
sources becomes harder toward fainter fluxes: the apparent photon
index in the 0.7--10 keV range changes from 2.1 at the flux of $\sim
10^{-11}$ to 1.6 at $\sim 10^{-13}$ \ergs\ (2--10 keV). This fact can
be explained by the rapid emergence of population with hard energy
spectra, as is clearly indicated in Figure~2(b). The evolution of
broad-band properties of sources solves the puzzle of discrepancy
discrepancy of the source counts between the soft (EMSS) and the hard
band (\ginga\ and \heao ). If we compare the \asca\ \logn s (including
Galactic objects) between above and below 2 keV, the hard band source
counts at $S\sim 10^{-13}$ \ergs\ (2--10 keV) matches the soft band
one when we assume a photon index of 1.6 for flux conversion, whereas
at brighter level of $S = 4\times 10^{-13} \sim 10^{-12}$ \ergs\
(2--10 keV), we have to use a photon index of about 1.9 to make them
match. The latter fact is consistent with the average 0.7--10 keV
spectrum at the same flux levels, and can be connected the the
``soft'' spectrum of the fluctuation observed with \ginga , which
shows a photon index of 1.8$\pm0.1$ in the 2--10 keV range (Butcher
\etal\ 1997).

The optical identification revealed that the major population at
fluxes of $10^{-13}$ \ergs\ are AGNs. The population of hard sources,
which are most responsible for making the average spectrum hard, are
X-ray absorbed sources. They are mostly identified as narrow line
(type-2) AGNs. The contribution of these type-2 AGNs is larger in the
hard band than in the soft band at the same flux limit. Recent results
of the 5--10 keV band survey by \sax\ confirms this tendency (Fiore
\etal\ 1999).  These results support the scenario that the CXB
consists of unabsorbed AGNs and absorbed AGNs, whose contribution
becomes more significant with decreasing fluxes and in harder energy
band.

We found, however, possible evidence that is not consistent with the ``unified
scheme'' of AGNs (e.g., Awaki \etal\ 1991), on which many AGN
synthesis models are based. The LSS results may imply deficiency of X-ray
luminous, absorbed AGNs, with $N_{\rm H} = 10^{22-23}$ at $z$ =
0.5--2, or in the X-ray luminosity range larger than $10^{44}$ \erg ,
although we cannot rule out possibility, for example, that there are
many luminous AGNs at $z>2$ with extreme heavy absorption of $N_{\rm
H} > 10^{24}$. On the other hand, there is another implication that
there could be a population of AGNs at high redshifts ($z>1$) that are
optically identified as type-1 AGNs but have apparently hard X-ray
spectra, although the origin of the hardness is not clear yet. Future
surveys by {\it Chandra} and {\it XMM} together with optical
identification of the AMSS sources will reveal the luminosity, number,
and spectral evolutions of extra-galactic populations including
absorbed AGNs, which will eventually lead us to full understanding of
the origin of the CXB.

\begin{figure}[h]
\centerline{\psfig{file=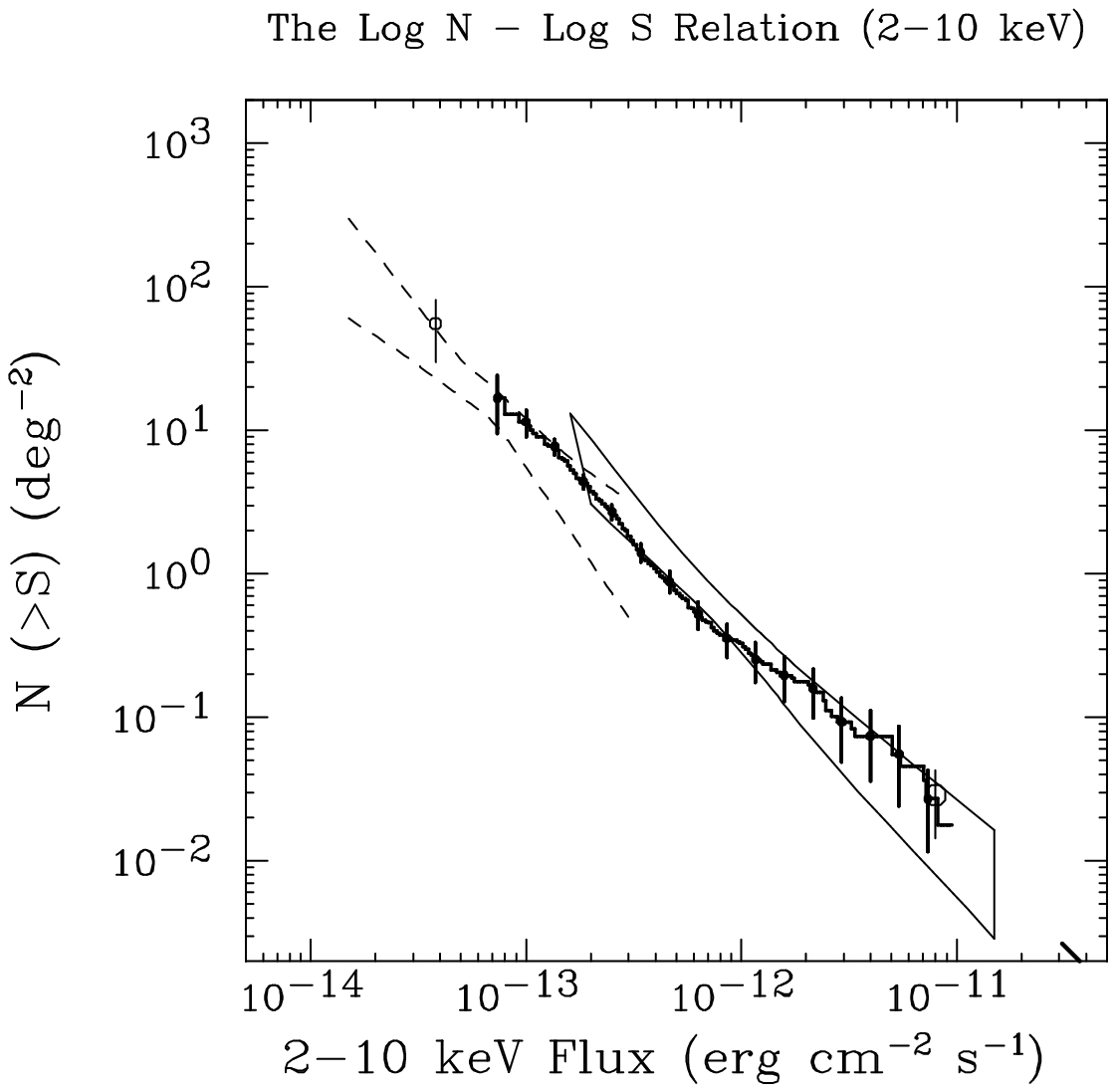, width=13cm}}
\caption[]{
Summary of the 2--10 keV \logn\ obtained by the \asca\ surveys,
compared with previous results. The steps are the combined results
from the LSS (Ueda \etal\ 1998) and the AMSS (Ueda \etal\ 1999b). The
faintest point at $4\times10^{-14}$ \ergs\ is derived from the DSS
utilizing the SIS data (Ogasaka \etal\ 1998). The trumpet shape
between two dashed lines indicates 1$\sigma$ error region from the
fluctuation analysis of \asca\ SIS deep fields (Gendreau, Barcons \&
Fabian 1998). The contour at $10^{-13}\sim10^{-11}$ \ergs\ represents
the constraints by the \ginga\ fluctuation analysis at 90\% confidence
level (Butcher \etal\ 1997). The open circle at $8\times10^{-12}$
\ergs\ corresponds to the source count by \ginga\ survey (Kondo
\etal\ 1991), and the thick-line above $3\times10^{-11}$ \ergs\ is the 
extragalactic \logn\ determined by \heao\ A2 (Piccinotti \etal\ 1982). 
All the horizontal bars represent 90\% statistical errors in source counts. 
}
\end{figure}

\begin{acknowledgements}

I thank all the collaborators of our \asca\ survey projects, especially,
M.~Akiyama, G.~Hasinger, H.~Inoue, Y.~Ishisaki, I.~Lehmann,
K.~Makishima, Y.~Ogasaka, T.~Ohashi, K.~Ohta, M.~Sakano, T.~Takahashi,
T.~Tsuru, W.~Voges, T.~Yamada, and A.~Yamashita.

\end{acknowledgements}

\setlength{\parindent}{0cm}

\end{document}